\documentclass[smallextended]{svjour3}       
\smartqed  
\usepackage[dvipsnames,usenames]{color}
\usepackage{graphicx}
 \usepackage{epsfig,float,lscape,ams,slashbox}
\begin{document}

\title{LRS Bianchi type I universes exhibiting Noether symmetry in the scalar-tensor Brans-Dicke
theory}

\titlerunning{Noether-symmetric LRS Bianchi type I universes in
scalar-tensor BD theory}

\author{Y. Kucukakca \and U. Camci \and \.{I}. Semiz }

\institute{ Y. Kucukakca \at Department of Physics, Akdeniz University, 07058 Antalya, Turkey \\
\email{ykucukakca@gmail.com} \and U. Camci \at Department of
Physics, Akdeniz University, 07058 Antalya, Turkey \\
\email{ucamci@akdeniz.edu.tr} \and \.{I}. Semiz \at Department of
Physics, Bogazici University, 34342 Bebek, Istanbul, Turkey \\
\email{semizibr@boun.edu.tr} }

\authorrunning{Y. Kucukakca, U. Camci, \.{I}. Semiz}

\journalname{General Relativity and Gravitation}

\date{Received: date / Accepted: date}

\maketitle


\begin{abstract}
Following up on hints of anisotropy in the cosmic microwave
background radiation (CMB) data, we investigate locally rotational
symmetric (LRS) Bianchi type I spacetimes with non-minimally
coupled scalar fields. To single out potentially more interesting
solutions, we search for Noether symmetry in this system. We then
specialize to the Brans-Dicke (BD) field in such a way that the
Lagrangian becomes degenerate (nontrivial) and solve the equations
for Noether symmetry and the potential that allows it. Then we
find the exact solutions of the equations of motion in terms of
three parameters and an arbitrary function. We illustrate with
families of examples designed to be generalizations of the
well-known power-expansion, exponential expansion and Big Rip
models in the Friedmann-Robertson-Walker (FRW) framework. The
solutions display surprising variation, a large subset of which
features late-time acceleration as is usually ascribed to dark
energy (phantom or quintensence), and is consistent with
observational data.

\end{abstract}

\section{Introduction} \label{int}
The recent decade has witnessed the observational evidence for the
cosmic acceleration of the universe, which has become the central
theme of modern cosmology. This evidence is built up of
observations of supernovae Type Ia  (SNe Ia) \cite{riess99},
cosmic microwave background radiation (CMB) \cite{netterfield02},
and large-scale structure of the Universe \cite{tegmark04}. In
Einstein's gravity, this acceleration cannot be explained by
normal matter or fields on Friedmann-Robertson-Walker (FRW) metric
background; therefore a mysterious cosmic fluid with negative
pressure, the so called \emph{dark energy}, is introduced.
Presently, understanding the nature of the dark energy is one of
the main problems in the research area of both theoretical physics
and cosmology. A simple example of dark energy is the cosmological
constant, equivalent to a fluid with the equation of state (EoS)
parameter $w=-1$, where $w=p/\rho$ in which $p$ is the pressure of
dark energy, and $\rho$ its energy density. However, the
cosmological constant model is subject to the so-called
\emph{fine-tuning} and \emph{coincidence problems}
\cite{copeland06}.

Dark energy, if it is a perfect fluid, must have EoS parameter
$w<-1/3$. If $-1/3 > w > -1$, the dark energy is called the
\emph{quintessence} \cite{cald98-sahni02}; if $ w < -1$, dark
energy is dubbed a \emph{phantom fluid} \cite{cald02}. In spite of
the fact that these models violate both the strong energy
condition $\rho + 3p > 0$ and the dominant energy condition $p +
\rho > 0$, and therefore may be physically considered undesirable,
the phantom energy is found to be compatible with current data
from SNe Ia observations, CMB anisotropy and the Sloan Digital Sky
Survey (SDSS) \cite{riess99,netterfield02,tegmark04}.

Dark energy behavior can be exhibited by possible new fundamental
fields acting on cosmological scales. The simplest such model is
the single component scalar field. Other alternatives in the
literature include the Cardassian expansion scenario
\cite{Freese}, the tachyon \cite{Sen00}, the quintom \cite{Li05},
the k-essence~\cite{Armen00} models and more.

Alternatively, modifications of Einstein's gravity have been
proposed to explain the cosmic acceleration of the universe; among
them $f(R)$ theories \cite{fr}. For example, quintessential
behavior of the parameter $w$ can be achieved in a geometrical way
in higher order theories of gravity \cite{cap002}. Scalar fields
with various couplings and potentials can be put in by hand, or
follow from the model naturally.

One of the simplest modification
of Einstein's gravity is the Brans-Dicke (BD) gravity theory
\cite{brans}, a well known example of a scalar-tensor theory which
represents the gravitational interaction using a scalar field in
addition to the metric field. This theory is parametrized by one
extra constant parameter, $W$. In the limit $W\rightarrow \infty$,
the BD theory reduces to Einstein's gravity \cite{fm}.
The conditions, where the dynamics of a self-interacting BD field
can account for the accelerated expansion, have been considered in
Ref.\cite{berto99}, where it was concluded that accelerated
expanding solutions can be obtained with a quadratic self-coupling
of the BD field and a negative equation of state (EoS) parameter.
Astrophysical data indicate that EoS parameter $w$ lies in an
interval of negative values roughly centered around $-1$.

The majority of popular cosmological models, including all the
ones referred to above, use the {\em cosmological principle}, that
is, they assume that the universe is homogeneous and isotropic. On
the other hand, there are hints in the CMB temperature anisotropy
studies that suggest that the assumption of statistical isotropy
is broken on the largest angular scales, leading to some
intriguing anomalies \cite{huterer}. To provide predictions for
the CMB anisotropies, one may consider the homogeneous but
anisotropic cosmologies known as Bianchi type spacetimes, which
include the isotropic and homogeneous FRW models \cite{barrow}. In
this study we consider the simplest of these, the locally
rotationally symmetric (LRS) Bianchi type I spacetime as an
anisotropic background universe model; note that this spacetime is
a generalization of flat ($k = 0$) FRW metric. Our aim is to
investigate the solutions of the field equations of scalar-tensor,
in particular, BD theory of gravity for the LRS Bianchi type I
spacetime using the Noether symmetry approach.

The Noether symmetry approach was introduced by De Ritis \emph{et
al.} \cite{ritis,demianski92} and Capozziello \emph{et al.}
\cite{capo93,capo00} to find preferred solutions of the field
equations and the dynamical conserved quantity. The Noether
theorem states that if the Lie derivative of a given Lagrangian
$L$ dragging along a vector field ${\bf X}$
 vanishes
\begin{equation}
\pounds_{\bf X} L = 0. \label{noether}
\end{equation}
then ${\bf X}$ is a symmetry for the dynamics, and it generates a
conserved current. Recently some exact solutions have been
presented in the scalar tensor theories following the Noether
symmetry approach that allows the potential to be chosen
dynamically, restricting the arbitrariness in a suitable way
\cite{ritis,demianski92,capo93,capo00,sanyal01,sanyal02,sanyal03,modak04,dabrowski,camci07,camci12,wei}.

    This paper is organized as follows. In the section \ref{general},
we present the field equations in scalar-tensor theory for the LRS
Bianchi type I spacetime. In section \ref{b1}, we search the
Noether symmetry of the Lagrangian of scalar-tensor theory for the
LRS Bianchi type I spacetime, and for the BD case, find new
variables that include the cyclic one. In section \ref{eos} we
derive the EoS of the BD field. In section \ref{fieldeqns}, we
give the solutions of the field equations by using new variables
obtained in section \ref{b1}. In section \ref{ornekler},  we
choose some examples corresponding to the well-known solutions in
the isotropic cases such as power law, exponentially expanding and
phantom/Big Rip models. We also make a conformal transform to the
so-called {\em Einstein frame} in Section \ref{einstein} and
discuss the problem in that context. Finally, in section
\ref{conc}, we conclude with a brief summary.

\section{The Lagrangian and the field equations}
\label{general}

The general form of the action that involves gravity non-minimally
coupled with a scalar field is given by \cite{sanyal03}, such that
\begin{eqnarray}
\mathcal{A} & = & \int{d^4 x \sqrt{-g} \left[ F(\Phi) R -
 \frac{W(\Phi)}{\Phi}\Phi_{c} \Phi^{c} - U(\Phi) \right]}.
\label{action}
\end{eqnarray}
Here $R$ is the Ricci scalar, $F(\Phi)$ and $W(\Phi)$ are generic
functions that describe the coupling, $U(\Phi)$ is the potential
for the scalar field $\Phi$, and $\Phi_a \equiv \Phi_{;a}$ stand
for the components of the gradient of the scalar field. Note that
we use Planck units. For $F(\Phi) = 1/2$ and $W(\Phi)= \Phi/2$,
the action reduces to the form of Einstein-Hilbert action
minimally coupled with a scalar field. The choice of $F(\Phi)$ and
$W(\Phi)$ give us other gravity theories such as  the BD theory,
for which $F(\Phi)=\Phi$ and $W(\Phi)=$constant.

Variation of the general form of the action with respect to metric
tensor yields the field equations
\begin{eqnarray}
& & F(\Phi) G_{ab}=T^{\Phi}_{ab} \label{feq}
\end{eqnarray}
where $G_{ab} = R_{ab} - \frac{1}{2} R g_{ab}$ is the Einstein
tensor,
\begin{eqnarray}
& & T^{\Phi}_{ab}= \frac{W}{\Phi}\Phi_{a}\Phi_{b} - \frac{W}{2
\Phi} g_{ab}\Phi_{c}\Phi^{c}- \frac{1}{2} g_{ab}U(\Phi)- g_{ab}
\Box F(\Phi) + F(\Phi)_{;ab} \qquad \label{emt}
\end{eqnarray}
is the energy-momentum tensor of scalar field, and $\Box$ is the
d'Alembert operator. It is clear that $T^{\Phi}_{ab}$ includes the
contributions from the non-minimal coupling and the scalar field
parts of the action varying with respect to metric. In
Ref.\cite{faraoni00}, it has been discussed that there are three
possible and inequivalent ways of writing the field equations,
corresponding to the ambiguity in the definition of the
energy-momentum tensor of non-minimally coupled scalar field. The
variation with respect to $\Phi$ gives rise to the generalized
Klein-Gordon equation governing the dynamics of the scalar field
\begin{eqnarray}
& & 2\frac{W}{\Phi} \Box \Phi + R
F'(\Phi)+\left(\frac{W'}{\Phi}-\frac{W}{\Phi^2}\right)\Phi_{c}\Phi^{c}
- U'(\Phi) = 0, \label{kg}
\end{eqnarray}
where the prime indicates the derivative with respect to $\Phi$.
Note that this equation is equivalent to the contracted Bianchi
identity. It follows from $G^{\,\,\,b}_{a\,;b} =0$ that using
Eq.(\ref{feq}) together with Eq. (\ref{emt}), yields
\begin{eqnarray}
& & \left[\frac{1}{F} T_a^{\,\,\,b (\Phi)} \right]_{;b} =0
\Leftrightarrow T^{\,\,\,b (\Phi)}_{a\,\,\,;b} = -
\frac{F_{;b}}{F} T_a^{\,\,\,b (\Phi)} \label{korunum}
\end{eqnarray}
From (\ref{emt}), we compute $T^{\,\,\,b (\Phi)}_{a\,\,\,;b}$ to
get
\begin{eqnarray}
& & T^{\,\,\,b (\Phi)}_{a\,\,\,;b} = \Phi_{a} \left[ \left(
\frac{W}{2\Phi} \right)' \Phi_c \Phi^c + \frac{W}{\Phi} \box \Phi
- \frac{1}{2} U'(\Phi) \right] + F_{;b} R_{a}^{\,\,b} \label{emt2}
\end{eqnarray}
Using Eq.(\ref{emt}), Eq.(\ref{emt2}) and $F_{;b} = F'(\Phi)
\Phi_b$ in Eq.(\ref{korunum}), we have
\begin{eqnarray}
& & \Phi_{a} \left[ 2\frac{W}{\Phi} \Box \Phi + R
F'(\Phi)+\left(\frac{W'}{\Phi}-\frac{W}{\Phi^2}\right)\Phi_{c}\Phi^{c}
- U'(\Phi)\right] = 0. \label{kg2}
\end{eqnarray}
This is obviously the generalized Klein-Gordon equation for
$\Phi_a \neq 0$.

The line element of the  LRS Bianchi type I spacetime has the form
\begin{equation} \label{metric-b1}
ds^2 = -dt^2 + A^2 dx^2 + B^2 \left( dy^2 + dz^2 \right),
\end{equation}
describing an anisotropic universe with equal expansion rate in
two of the three dimensions. The Ricci scalar of this spacetime is
\begin{equation}
R = 2 \left[ \frac{\ddot{A}}{A} + 2 \frac{\ddot{B}}{B}+
\frac{\dot{B}^2}{B^2} + 2\frac{\dot{A}\dot{B}}{A B}  \right],
\end{equation}
where the dot represents differentiation with respect to $t$. For
the metric (\ref{metric-b1}), the field and generalized
Klein-Gordon equations can be obtained from Eqs.(\ref{feq}) and
(\ref{kg}) respectively
\begin{eqnarray}
& & \frac{\dot{B}^2}{B^2} + 2\frac{\dot{A} \dot{B}}{A B} +
\frac{F'}{F} \left( \frac{\dot{A}}{A} + 2 \frac{\dot{B}}{B}
\right) \dot{\Phi} - \frac{1}{2 F}\left[
\frac{W}{\Phi}\dot{\Phi}^2+ U \right] =0, \label{feq1}
\\& &  2 \frac{\ddot{B}}{B} + \frac{\dot{B}^2}{B^2}
 + \frac{F'}{F} \left[ \ddot{\Phi} + 2 \frac{\dot{B}}{B} \dot{\Phi}
\right] + \frac{1}{2 F}\left( \frac{W}{\Phi}+2F'' \right)
\dot{\Phi}^2 - \frac{U}{2 F} = 0, \quad \label{feq2}
\\& &  \frac{\ddot{A}}{A} + \frac{\ddot{B}}{B}+ \frac{\dot{A}
\dot{B}}{A B} + \frac{F'}{F} \left[ \ddot{\Phi} + \left(
\frac{\dot{A}}{A} + \frac{\dot{B}}{B}\right) \dot{\Phi} \right]
\nonumber  \\& & \qquad \quad + \frac{1}{2 F}\left( \frac{W}{\Phi}
+ 2F'' \right) \dot{\Phi}^2 - \frac{U}{2 F}  = 0, \label{feq3}
\\& &  \frac{\ddot{A}}{A}+ 2\frac{\ddot{B}}{B} +
\frac{\dot{B}^2}{B^2} + 2\frac{\dot{A}\dot{B}}{A B} -
\frac{W}{F'\Phi}\left[ \ddot{\Phi} + \left( \frac{\dot{A}}{A} +
2\frac{\dot{B}}{B}\right) \dot{\Phi}\right] \nonumber  \\& &
\qquad \quad -\frac{\dot{\Phi}^2}{2F'}\left(\frac{W'}{\Phi}
-\frac{W}{\Phi^2}\right)-\frac{U'}{2F'} = 0 \label{feq4}
\end{eqnarray}
where $F' \neq 0$. The Lagrangian density of the LRS Bianchi type
I spacetime is
\begin{eqnarray} \label{lag-b13ks}
& &  L = -2  F A \dot{B}^2 - 4  F B \dot{A}\dot{B} - 2  F' B^2
\dot{A} \dot{\Phi} - 4 F' A B \dot{B} \dot{\Phi} \nonumber  \\& &
\qquad \quad + A B^2 \left[ \frac{W\dot{\Phi}^2}{\Phi} - U(\Phi)
\right]. \label{lagra}
\end{eqnarray}
Using such a Lagrangian, one may obtain the Euler-Lagrange
equations as given in (\ref{feq2})-(\ref{feq4}). The \emph{energy
function}, $E_{L}$, associated with the Lagrangian
(\ref{lag-b13ks}) is found as
\begin{eqnarray}
E_{L} &=& \frac{\partial L}{\partial \dot{A}} \dot{A} +
\frac{\partial L}{\partial \dot{B}} \dot{B} +
\frac{\partial L}{\partial \dot{\Phi}} \dot{\Phi} - L \nonumber \\
&=& \frac{\dot{B}^2}{B^2} + 2\frac{\dot{A} \dot{B}}{A B} +
 \frac{F'}{F} \left( \frac{\dot{A}}{A} + 2
\frac{\dot{B}}{B} \right) \dot{\Phi} - \frac{1}{2 F}\left[
\frac{W}{\Phi}\dot{\Phi}^2+ U \right]. \label{e-b13ks}
\end{eqnarray}
Therefore, it is obvious that the (0,0)-field equation given by
(\ref{feq1}) is equivalent to $E_{L} = 0$.

\section{Noether symmetry approach} \label{b1}

In this section we seek the Noether symmetry of the the Lagrangian
(\ref{lagra}). The configuration space of this Lagrangian is $Q =
(A, B, \Phi)$, whose tangent space is $TQ
=(A,B,\Phi,\dot{A},\dot{B},\dot{\Phi})$. The existence of Noether
symmetry implies the existence of a vector field ${\bf X}$ such
that
\begin{equation}
{\bf X} = \alpha \frac{\partial}{\partial A} + \beta
\frac{\partial}{\partial B} + \gamma \frac{\partial}{\partial
\Phi} + \dot{\alpha} \frac{\partial}{\partial \dot{A}} +
\dot{\beta} \frac{\partial}{\partial \dot{B}} + \dot{\gamma}
\frac{\partial}{\partial \dot{\Phi}} \label{vec1}
\end{equation}
where $\alpha, \beta$ and $\gamma$ are depend on $A, B$ and
$\Phi$. Hence the Noether equation given by (\ref{noether}) yields
the following set of equations
\begin{eqnarray}
& &  2 \frac{\partial \beta}{\partial A} + B \frac{F'}{F}
\frac{\partial \gamma}{\partial A} =0, \label{neq1} \\& &
\frac{\alpha}{2} + B \frac{\partial \alpha}{\partial B} + A
\frac{\partial \beta}{\partial B} + A \frac{F'}{F} \left(
\frac{\gamma}{2} + B \frac{\partial \gamma}{\partial B} \right) =
0, \label{neq2} \\& &  \frac{W}{\Phi}\left( \frac{\alpha}{2} +
\beta \frac{A}{B}+ A \frac{\partial \gamma}{\partial \Phi}\right)+
\frac{A}{2}\left(\frac{W'}{\Phi}-\frac{W}{\Phi^2}\right)\gamma
\nonumber \\& & \qquad \quad  - F' \left( \frac{\partial
\alpha}{\partial \Phi} + 2 \frac{A}{B} \frac{\partial
\beta}{\partial \Phi} \right) = 0, \label{neq3} \\& &  \beta + B
\frac{\partial \alpha}{\partial A} + A \frac{\partial
\beta}{\partial A} + B \frac{\partial \beta}{\partial B} + B
\frac{F'}{F} \left( \gamma + A \frac{\partial \gamma}{\partial A}
+ \frac{B}{2} \frac{\partial \gamma}{\partial B} \right) = 0,
\label{neq4} \\& &  2 \frac{\partial \beta}{\partial \Phi} +
\frac{F'}{F} \left( 2 \beta + B \frac{\partial \alpha}{\partial A}
+ B \frac{\partial \gamma}{\partial \Phi} + 2 A \frac{\partial
\beta}{\partial A} \right) \nonumber
\\& & \qquad \quad  + \frac{F''}{F} B \gamma - \frac{W}{
F\Phi} A B \frac{\partial \gamma}{\partial A} = 0, \label{neq5}
\\& &  \frac{\partial \alpha}{\partial \Phi} + \frac{A}{B}
\frac{\partial \beta}{\partial \Phi} +  \frac{F'}{F} \left( \alpha
+ \frac{A}{B} \beta + \frac{B}{2} \frac{\partial \alpha}{\partial
B} + A \frac{\partial \beta}{\partial B} + A \frac{\partial
\gamma}{\partial \Phi} \right) \nonumber
\\& & \qquad \quad + \frac{F''}{F} A \gamma - \frac{W \, A B}{2 F \Phi} \frac{\partial
\gamma}{\partial B} = 0, \label{neq6}
\\& &  (B \alpha + 2 A \beta ) U + A\, B \, \gamma U' = 0.
\label{neq7}
\end{eqnarray}
Leaving these equations as reference for future work, we now
specialize to the Brans-Dicke case, where $F = \Phi$ and $W$ is
constant. We also require the Hessian determinant, $D = \Sigma
\left| \frac{\partial^2 L}{\partial \dot{Q}_i \partial \dot{Q}_j}
\right|$, to vanish in order to get nontrivial solutions. This
condition reads for our case as
\begin{equation} \label{hessian}
D = -\frac{16 A B^4 F}{\Phi} (3 \Phi F'^2 + 2 W F) = 0,
\end{equation}
which also determines a $W$ value. So we will be working with the
coupling functions
\begin{equation}
F = \Phi, \qquad W=-\frac{3}{2}.   \label{f-degenerate}
\end{equation}
Thus the Lagrangian (\ref{lagra}) becomes degenerate, and the BD
action has the form
\begin{eqnarray}
\mathcal{A_{BD}} & = & \int{d^4 x \sqrt{-g} \left[ \Phi R +
 \frac{3}{2 \Phi}\Phi_{c} \Phi^{c} - U(\Phi) \right]}
\label{action-bd}
\end{eqnarray}
which can be easily related the conformal relativity by defining
new scalar field $\varphi$ as $\Phi = \varphi^2 / 12$,
transforming the action into
\begin{eqnarray}
\mathcal{A_{BD}} & = & \frac{1}{2} \int{d^4 x \sqrt{-g} \left[
\frac{1}{6}\varphi^2 R + \varphi_{c} \varphi^{c} - U(\varphi)
\right]}. \label{action-bd-confinv}
\end{eqnarray}
This action is of course conformally invariant, since the
application of the conformal transformation formulas together with
the appropriate integration of the boundary term gives the same
form of this action (see Refs. \cite{fm} and \cite{dabrowski}).

Dabrowski \emph{et al.} \cite{dabrowski} have shown that the
anisotropic non-zero spatial curvature models of Bianchi  types I,
III and Kantowski-Sachs type are admissible in $W =-3/2$ BD
theory. They have given solutions of BD field equations for these
spacetimes without the potential $U(\Phi)$. In a previous work
\cite{camci07}, we studied the case $F = \frac{\epsilon}{12}(\Phi
- \Phi_0)^2 ,W=\frac{\Phi}{2}$ for the Bianchi types I and III,
and Kantowski-Sachs spacetimes, where $\epsilon$ is a parameter
depending the signature of metric, but did not study the
qualitative (acceleration etc.) behaviour of the universes in the
solutions found.

In our present study we find the solutions of BD field equations
with potential $U(\Phi)$ for the LRS Bianchi type I spacetime,
although we derive the equations for a general function $W(\Phi)$.
In this context, the solutions of the above set of differential
equations (\ref{neq1})-(\ref{neq7}) for $\alpha, \beta$,  $\gamma$
and potential $U(\Phi)$ are obtained as
\begin{eqnarray}
& & \alpha = \left(AB \Phi^{3/2}\right)^{-1}, \,\,  \beta =
\frac{1}{2} \left(A^2 \Phi^{3/2}\right)^{-1}, \,\, \gamma = -
\left(A^2 B \Phi^{1/2}\right)^{-1}, \nonumber \\& &  U(\Phi) =
\lambda \Phi^2 \label{pot}
\end{eqnarray}
where $\lambda$ is a constant. The potential found indicates that
for the Noether symmetry to be present, the scalar field must be a
(massive) {\em free} field. The vector field ${\bf X}$ generating
the Noether symmetry and determining the dynamics of LRS Bianchi
type I metric is given by
\begin{eqnarray}
{\bf X} =\frac{1}{AB \Phi^{3/2}} \Big{[}
\frac{\partial}{\partial{A}}+
\frac{B}{A}\frac{\partial}{\partial{B}}-
\frac{\Phi}{A}\frac{\partial}{\partial{\Phi}}-
\left(\frac{\dot{A}}{A}+\frac{\dot{B}}{B}+\frac{3\dot{\Phi}}{2\Phi}\right)
\frac{\partial}{\partial{\dot{A}}} \nonumber
\\ \qquad \qquad \qquad
-\left(\frac{\dot{A}}{A}+\frac{3\dot{\Phi}}{4\Phi}\right)
\frac{\partial}{\partial{\dot{B}}} +
\left(\frac{2\dot{A}}{A}+\frac{\dot{B}}{B}+\frac{1\dot{\Phi}}{2\Phi}\right)
\frac{\partial}{\partial{\dot{\Phi}}}\Big{]}\label{vec2}
\end{eqnarray}

\section{Equation of state } \label{eos}

Using $F(\Phi)= \Phi$ and $W=-3/2$ in Eq. (\ref{emt}) we can
derive the energy density of the scalar field
\begin{eqnarray}
& & \rho^\Phi =-\frac{3}{4}\frac{\dot{\Phi}^2}{\Phi}+
\frac{U(\Phi)}{2}-3H \dot{\Phi} \label{en-dens}
\end{eqnarray}
and the directional pressure due to the scalar field
\begin{eqnarray}
& & p_i^\Phi =-\frac{3}{4}\frac{\dot{\Phi}^2}{\Phi}-
\frac{U(\Phi)}{2} + \ddot{\Phi} +\left(3H - H_i\right) \dot{\Phi},
\quad i = x, y, z \label{pres}
\end{eqnarray}
for the $x,y$ and $z$ directions. Here $H_i$ represents the
directional Hubble parameters in the directions of $x,y$ and $z$
respectively, and may be defined as
\begin{eqnarray}
& & H_x = \frac{\dot{A}}{A}, \quad H_y = H_z = \frac{\dot{B}}{B}.
\label{Hubble-D1}
\end{eqnarray}
The mean Hubble parameter for LRS Bianchi type I metric is given
by
\begin{eqnarray}
& & H = \frac{\dot{a}}{a}=\frac{1}{3} \left(H_x + 2H_y \right),
\label{meanHubble}
\end{eqnarray}
where we define $a = (AB^2)^{1/3}$ as the average scale factor of
the Universe. Also of interest is the volumetric deceleration
parameter (for the isotropic case, simply the "deceleration
parameter"), defined by
\begin{equation}
q=-\frac{\ddot{a}a}{\dot{a}^2} \equiv -1 - \frac{\dot{H}}{H^2}
\label{decparam}
\end{equation}
so that it is dimensionless.

The pressure is a vectorial quantity, as would be expected for
anisotropic expansion, and thus the EoS parameter of the scalar
field for the LRS Bianchi type I spacetime may be determined
separately on each spatial axis: $w_i^{\Phi} (t)=
p_i^{\Phi}/\rho^{\Phi}$, i.e.
\begin{eqnarray}
& & w_i^{\Phi} = \frac{p_i^{\Phi}}{\rho^\Phi}
=\frac{\frac{3}{4}\frac{\dot{\Phi}^2}{\Phi} + \frac{U(\Phi)}{2} -
\ddot{\Phi} - \left(3H-H_i\right)
\dot{\Phi}}{\frac{3}{4}\frac{\dot{\Phi}^2}{\Phi} -
\frac{U(\Phi)}{2} + 3H \dot{\Phi}} \label{eos-1}
\end{eqnarray}
and the average EoS parameter of scalar field for the LRS Bianchi
type I spacetime may be defined as
\begin{eqnarray}
& & w^{\Phi}=\frac{p^\Phi}{\rho^\Phi}=\frac{1}{3} \left(w_x + 2
w_y \right) =\frac{\frac{3}{4}\frac{\dot{\Phi}^2}{\Phi} +
\frac{U(\Phi)}{2} - \ddot{\Phi} - 2 H
\dot{\Phi}}{\frac{3}{4}\frac{\dot{\Phi}^2}{\Phi} -
\frac{U(\Phi)}{2} + 3H \dot{\Phi}} \label{a-eos}
\end{eqnarray}
where the $p^{\Phi}$ is the isotropic pressure \cite{Calogero}.

In the anisotropic case, $w$ does not directly determine the sign
of $q$ as it does in the well-known (spatially) flat FRW
cosmologies. To understand this, consider the Raychaudhuri
Equation~\cite{gron}
\begin{equation}
\dot{\theta} + \frac{1}{3}\theta^{2} +
\sigma_{\mu\nu}\sigma^{\mu\nu} -  \omega_{\mu\nu}\omega^{\mu\nu} +
\frac{\kappa}{2}(\rho+3p) - \Lambda = 0  \label{raychaudhuri}
\end{equation}
where $\theta$ is the {\em expansion scalar}, $\sigma_{\mu\nu}$ is
the {\em shear tensor}, $\omega_{\mu\nu}$ is the {\em vorticity
tensor} (descriptions/definitions also given in~\cite{gron}) and
we take $\Lambda=0$. If comoving particles in a metric move
geodesically, the expansion scalar also gives the expansion rate
of the space itself, when we apply this equation to a swarm of
such particles. This is the case for the LRS Bianchi I metric,
where  $\theta$ turns out to be equal to $3H$, therefore
$\dot{\theta} + \frac{1}{3}\theta^{2}$ is $-3H^{2}q$. The
vorticity tensor vanishes for this case, and
\begin{equation}
\sigma_{\mu\nu}\sigma^{\mu\nu} = \frac{2B(t)^{2}}{3A(t)^{2}}
\left[\frac{d}{dt}\left(\frac{A(t)}{B(t)}\right)\right]^{2}.
\label{shear}
\end{equation}
Therefore in the isotropic case $\sigma_{\mu\nu}\sigma^{\mu\nu}$
also vanishes and since in that case $\rho \propto H^{2}$, we find
that $q$ is proportional to $1+3w$. This correspondence is
obviously broken when there is anisotropy, and therefore $w$ is
not that useful a parameter.

\section{The solutions from new coordinates and Lagrangian}
\label{fieldeqns}
The solutions of dynamical Eqs.(\ref{feq1})-(\ref{feq4}) are not
easy to evaluate in the present form. In order to simplify these
equations we can search for the cyclic variable(s). In case of Noether symmetry,
we should introduce new variables instead of old variables, i.e. a point transformation $\{A, B,
\Phi\}\rightarrow \{\mu, \nu, u \}$ in which it is assumed that $\mu$
is the cyclic coordinate. A general discussion of this procedure has been given in Ref.\cite{capo96}. The new variables $\{\mu, \nu, u \}$ satisfy the
following equations
\begin{eqnarray} & & i_{\bf X} d\mu =1, \quad i_{\bf X} d\nu =0,
\quad i_{\bf X} du =0 \label{car1}
\end{eqnarray}
where  $i_{\bf X}$ is the interior product operator of ${\bf X}$.
Using Eq.(\ref{pot}),  a solution of the above Eqs.(\ref{car1})
yield
 \begin{eqnarray}
\mu = A^2 B \Phi^{3/2}, \quad \nu = B \Phi^{1/2}, \quad u = A \Phi.
\label{trns}
\end{eqnarray}
The inverse transformation of these variables are
\begin{eqnarray}
A = \frac{\mu}{\nu u}, \quad B =\frac{(\mu \nu)^{1/2}}{u},
\quad \Phi =\frac{\nu u^2}{\mu}, \label{i-trns}
\end{eqnarray}
from which it follows that
\begin{eqnarray}
a (t) = \frac{\mu^{2/3}}{u}, \label{a-new}
\end{eqnarray}
for the average scale factor of the Universe defined by $a= \left(
A B^2 \right)^{1/3}$. Considering the transformation  of
variables, the coupling functions (\ref{f-degenerate}) and the
potential $U(\Phi)$ given in (\ref{pot}), the Lagrangian
(\ref{lagra}) becomes
\begin{eqnarray}
L = - 2 u^{-1} \dot{\mu}\dot{\nu}-
\lambda u \nu^2  \label{lag2}
\end{eqnarray}
which does not depend on $\mu$ (i.e.
$\frac{\partial{L}}{\partial{\mu}}=0$), as desired. This
independence is preserved if we make further transformations in
the $u-\nu$ plane; and also is inherent in the procedure described
in Ref.\cite{capo96}. We used this freedom to get as simple a
Lagrangian as possible.

The Euler-Lagrange equations relative to this Lagrangian are
\begin{eqnarray}
& & \dot{\nu}- l_0 u =0, \label{EL} \\& & \dot{\mu} \dot{\nu} -
\frac{\lambda}{2} \nu^2 u^2 =0, \label{E-L1} \\& &
\frac{\ddot{\mu}}{\mu} -\frac{\dot{\mu} \dot{u}}{\mu u} - \lambda
\frac{u^2 \nu}{\mu} =0, \label{E-L2}
\end{eqnarray}
where $l_0$ is a constant of motion associated with the coordinate
$\mu$.  Here $i_{\bf X}{\Theta_L} = a_0 = -2 l_0$ and $\Theta_L$
is the Cartan one-form, and Eq.(\ref{E-L1}) is equivalent to the
vanishing of the energy functional, as noted before.

By substituting $u$ from Eq.(\ref{EL}) into Eq.(\ref{E-L1}) we get
\begin{equation}
\mu(t)=l_1 \nu^3 + l_2,  \label{sol1}
\end{equation}
where $l_1 = \lambda/(6 l_0^2), \, l_0 \neq 0$, and $l_2$ is an
integration constant. The remaining equation (\ref{E-L2}) is
identically satisfied now. Inserting (\ref{EL}) and (\ref{sol1})
into (\ref{i-trns}),  we obtain the metric functions and scalar
field in terms of the arbitrarily specifiable $\nu$ (henceforth to be called the {\em seed function}), and $u = \dot{\nu}/l_0$ as follows:
\begin{eqnarray}
& & A(t) = l_{0} \frac{ l_{1} \nu(t){^3} + l_{2} }{\nu(t)
\dot{\nu}(t)}, \label{metric-A1}
\\& & B(t) = \frac{l_0}{\dot{\nu}(t)} \sqrt{\nu(t)\left(l_{1} \nu(t){^3} + l_{2}\right)}, \label{metric-B1}
\\& & \Phi(t) = \frac{\nu(t)  \dot{\nu}(t)^{2}}{l_0^{2} \left(l_{1} \nu(t){^3} + l_{2}\right)}. \label{sclr-fld}
\end{eqnarray}
Thus, using (\ref{metric-A1}) and (\ref{metric-B1}), the LRS
Bianchi type I metric (\ref{metric-b1}) becomes
\begin{equation} \label{metric-son}
ds^2 = - dt^2 + l_0^2 \left( l_1 \nu(t)^3 + l_2 \right) \frac{\nu(t)}{\dot{\nu}(t)^2} \left[ \left( l_1 + \frac{l_2}{\nu(t)^3} \right) dx^2 + dy^2 + dz^2 \right].
\end{equation}
Then, the average scale factor of the Universe takes the form
\begin{eqnarray}
& & a(t) = \frac{l_0}{\dot{\nu}(t)} \left(l_{1} \nu(t){^3} +
l_{2}\right)^{2/3}.\label{scale-f1}
\end{eqnarray}
Also the case of isotropy can be seen, by requiring $A\propto B$,
to correspond to $l_2=0$.

In the general anisotropic case, the directional Hubble parameter
in $x$ direction defined in Eq.(\ref{Hubble-D1}) is found as
\begin{eqnarray}
& & H_x =
\frac{\dot{\mu}}{\mu}-\frac{\dot{\nu}}{\nu}-\frac{\dot{u}}{u}=-\frac{\ddot{\nu}}{\dot{\nu}}+
\frac{\dot{\nu}}{\nu} \left[ \frac{3 l_{1} \nu^{3}}{l_{1} \nu{^3}
+ l_{2}} - 1 \right], \label{Hubble-genel-x}
\end{eqnarray}
the other directional Hubble parameters in the directions $y$ and
$z$ are similarly found as
\begin{eqnarray}  & & H_y = H_z = \frac{1}{2}
\left(\frac{\dot{\mu}}{\mu}+\frac{\dot{\nu}}{\nu}
\right)-\frac{\dot{u}}{u}= -\frac{\ddot{\nu}}{\dot{\nu}}+
\frac{\dot{\nu}}{2\nu} \left[ \frac{3 l_{1} \nu^{3}}{l_{1} \nu{^3}
+ l_{2}} + 1 \right], \label{Hubble-genel-y}
\end{eqnarray}
and the mean Hubble parameter defined in Eq.(\ref{meanHubble}) is
found as
\begin{eqnarray}
& & H = \frac{2}{3}
\frac{\dot{\mu}}{\mu}-\frac{\dot{u}}{u}=-\frac{\ddot{\nu}}{\dot{\nu}}+
\frac{ 2 l_{1} \nu^{2} \dot{\nu}}{l_{1} \nu{^3} + l_{2}}.
\label{meanHubble-genel}
\end{eqnarray}

At this point, we would like to emphasize that particular
solutions can be found, that is, the scale factors $A(t), B(t)$,
the potential $\Phi(t)$ and the average scale factor $a(t)$ can be
obtained by specifying the function $\nu(t)$. In the next section,
we will find such examples.

\section{Examples}
\label{ornekler}

To illustrate our family of solutions for the dynamics of LRS
Bianchi type I cosmologies containing a scalar field, we choose
examples corresponding to  the well-known solutions in the
isotropic case, that is, power-law models, i.e. $a_{\rm isotropic}
\propto  t^{\eta}$, exponentially expanding models, i.e. $a_{\rm
isotropic} \propto  e^{\chi t}$, and phantom/Big Rip models,  i.e.
$a_{\rm isotropic} \propto (t_{c}-t)^{-\sigma}$, where $\eta,
\chi$ and $\sigma$ are positive constants. We can find the form of
each $\nu(t)$ by solving from Eq.(\ref{scale-f1}) after setting
$l_{2}=0$.

The most general form we find for the power-law models is
$\nu(t)=\frac{C_{1}}{C_{2}-t^{\eta-1}}$, but for simplicity we use
$\nu(t)=a_{1} t^n$. For the exponential case, the most general
form is $\nu(t)=\frac{C_{3}}{C_{4}-e^{-\chi t}}$, but we use
$\nu(t)=a_{2} e^{kt}$.  Similarly, for the phantom/Big Rip models,
the most general form is
$\nu(t)=\frac{C_{5}}{C_{6}-(t_{c}-t)^{\sigma+1}}$, but we use
$\nu(t)=\frac{a_{3}}{(t_{c}-t)^{m}}$ ($n$, $k$ or $m$ are not necessarily integer).

\subsection{Power-law seed}
\label{powerlaw}

When we use $\nu(t)= a_1 t^{n}$, where $a_1$ and $n$ are nonzero
constants, the scale factors read
\begin{equation}
A(t) = \frac{l_0}{a_{1}^{2} n} t^{1-2n} (l_1 a_1^3 t^{3n}+l_2), \;\;
 B(t)= \frac{l_0}{n \sqrt{a_{1}}} t^{1-n/2}  \left[ l_1 a_1^3 t^{3n} + l_2 \right]^{1/2},
\label{powerlaw-A-B}
\end{equation}
the average scale factor is found from Eq.(\ref{scale-f1}) as
\begin{eqnarray}
& &  a(t) =   \frac{l_0 }{a_1 n} \left[ l_1 a_1^3 t^{3n} + l_2
\right]^{2/3} t^{1-n}, \label{scale-f3}
\end{eqnarray}
the scalar field from  Eq.(\ref{sclr-fld}) as
\begin{eqnarray}
& & \Phi(t) = \frac{ a_1^3 n^2 t^{3n-2}}{l_0^2 \left(l_1 a_1^3
t^{3n} + l_2  \right)}, \label{sclr-fld3}
\end{eqnarray}
and the mean Hubble parameter from Eq.(\ref{meanHubble}) as
\begin{eqnarray}
H = \frac{2 l_1 a_1^3 n t^{3n -1}}{ l_1 a_1^3 t^{3n} +l_2 } +
\frac{1-n}{t}. \label{meanHubble3}
\end{eqnarray}

Inspection of the scale factors $A(t)$, $B(t)$ and $a(t)$ shows
that if $l_{1}$ or $l_{2}$ is zero, they will reduce to single
powers of $t$. But also in the general case where both $l_{1}$ and
$l_{2}$ are nonzero, the scale factors will approach one power of
$t$ near zero and another near infinities. These powers depends on
the sign of $n$: For negative $n$, near $t=0$ we
have\label{powerScaleLimits}  $A(t), B(t), a(t) \propto t^{1+n}$,
near infinities we have $A(t) \propto t^{1-2n}$, $B(t) \propto
t^{1-n/2}$ and $a(t) \propto t^{1-n}$; for positive $n$, the
behaviors near zero and infinities are interchanged. Therefore, we
need to treat the cases of vanishing $l_{1}$ or $l_{2}$ separately
from the general case. These cases will be split into subcases,
for example, according to the value of $n$, etc. Each line in the
Tables \ref{tb:l2zero},\ref{tb:l1zero} and \ref{tb:l1l2nonzero}
shows the timeline of  an eventual subcase from $t \rightarrow
-\infty$ to $t \rightarrow +\infty$, which usually describes {\em
more than one} universe, since the timeline may be interrupted by
singularities,  shown by expressions in square brackets in the
lines of the tables. Since solutions cannot be taken valid {\em
through} singularities; each interval between singularities should
be considered an independent solution/universe. The universes not
extending to the right or left edge of a line have finite
lifetime.

Obvious candidates for singularities are $t=0$ and $t \rightarrow
\pm \infty$, in light of the above discussion. Moreover, the $(
l_1 a_{1}^{3}t^{3n}+l_2)$ terms in the scale factors can vanish at
some finite $t$ value $t_{1}$, or diverge at  $t=0$, if both
$l_{1}$ and $l_{2}$ are nonzero. The meaning of a singularity not
only depends on its mathematical behavior, but also on its
relative time-relation to the observers:  If observers see (or
calculate) vanishing comoving volume in their finite past or
finite future, they will call that singularity a Big Bang (BB) or
a Big Crunch (BC); This is shown as "BC[0]BB" in the tables.
Similarly, observers calculating diverging of the comoving volume
in their finite future will call that singularity a Big Rip (BR),
and the case of diverging comoving volume in the finite past we
will call it an Inverse Big Rip (iBR).

The square brackets should contain an ordered triple, where the
first entry shows the behavior of $A(t)$, the second entry the
behavior of $B(t)$ and the third, the behavior of $a(t)$. The
behavior is indicated by the symbols $0, C$ or $\infty$ to denote
that the relevant scale factor vanishes/goes to a finite
number/diverges at that point. To make the tables more compact, an
ordered pair (describing the behavior of $A(t)$ and $B(t)$;
whenever these make the behavior of $a(t)$ obvious) or a single
entry (indicating that the behavior of all three scale factors is
the same) may be used. Finally, one should also note that negative
$t$ can only be considered if $n$ is a rational number with an odd
denominator, not for general $n$. \label{negNrest}

We discuss the three above-mentioned cases in order of increasing
complexity: First, we discuss the isotropic $l_{2}=0$ case, then
the simple anisotropic $l_{1}=0$ case, and finally the general
anisotropic case where both $l_{1}$ and $l_{2}$ are nonzero.

\subsubsection{The $l_{2}=0$ (isotropic) case.}

As stated after Eq.(\ref{scale-f1}), in this case the LRS Bianchi
type I spacetime becomes isotropic, i.e. reduces to the well-known
spatially flat FRW metric. The scale factors become proportional
to each other, and
\begin{eqnarray}
& & a(t) = a_0 t^{1+n} \label{scale-f4}
\end{eqnarray}
where $a_0 = a_1 \frac{l_0 l_1^{2/3}}{n}$. This means that $a(t)$
can vanish or diverge at zero or infinity, and the critical $n$
values are -1 and 0, leading to the subcases shown in Table
\ref{tb:l2zero}. The cases are

\begin{table}[htbp!]
\centering
\begin{tabular}{|c|lcccr|}
\hline
\backslashbox{$n$}{$t$}  & $-\infty$ &   & $0$ &  & $+\infty$ \\ \hline

$n < -1$     & [0]Z    & & BR[$\infty$]iBR     & & Z[0] \\ \hline

$n = -1$     & (C) &---static--- & (C) &---static--- & (C) \\ \hline

$-1 < n < 0$ & [$\infty$]I & "decelerating" & BC[0]BB & decelerating & I[$\infty$] \\ \hline

$n \rightarrow 0$       & [$\infty$]I        & linear & BC[0]BB
 & linear & I[$\infty$] \\ \hline

$n > 0$   & [$\infty$]I    & "accelerating" & BC[0]BB & accelerating & I[$\infty$] \\ \hline

\end{tabular}
\caption{Universes derived for  $l_{2} = 0$ ({\it isotropic}
universes) where $\nu(t)$ is a power of $t$. Each line shows a
timeline from $-\infty$ to $\infty$, but may represent {\em
multiple} universes delimited by singularities, shown by square
brackets. The expressions in the brackets refer to the scale
factor $a(t)$. Also, I: infinite scale factor at infinite future
or past, BR: Big Rip, i: Inverse, BC: Big Crunch, BB: Big Bang,
Z: vanishing scale factor at infinite future or past [Also see
text and the note about negative time on page \pageref{negNrest}].
\label{tb:l2zero}}
\end{table}

\begin{itemize} 

\item $n < -1$: Negative times\footnote{subject to the condition
on $n$ mentioned on page \pageref{negNrest}} describe a universe
that expands from a vanishingly small scale factor in the infinite
past, and then the scale factor diverges within a finite time: one
might call this a Zero-Big Rip (Z-BR) universe. Positive times
describe another universe, whose behavior is the time-reverse of
the first, an iBR-Z universe.

\item $n = -1$: This universe, being static and flat, is
equivalent to Minkowski space.

\item $-1 < n < 0$: For positive times, this case represents
decelerating (spatially) flat FRW universes, including the
radiation-dominated ($n=-1/2$) and matter-dominated ($n=-1/3$)
expansions, in particular. For negative times$^1$, however, the
universes contract from infinite scale factor in the infinite past
to a Big Crunch at $t=0$; so the universes represented on the last
three lines of Table \ref{tb:l2zero} may be called Infinity-Big
Crunch (I-BC) and Big Bang-infinity (BB-I) universes. The I-BC
universes are decelerating, too: The negative $\dot{a}(t)$ [for
positive $a(t)$] becomes more negative with time.

\item $n \rightarrow 0$: This case can be analyzed with the help
of an infinite scale transformation [note that Eqs.
(\ref{EL})-(\ref{E-L2}) are satisfied for constant $\nu(t)$], and
describes universes linearly expanding (positive time) or
contracting (negative time$^1$) with increasing time.

\item $n > 0$: Positive times describe a universe that starts with
a Big Bang and expands forever with increasing speed; negative
times$^1$ describe a universe with time-reversed behavior.

\end{itemize}

As is well-known (and mentioned above), in FRW cosmology the
behavior of the universe and the properties of the (possibly
efective) fluid contained in the universe are related. In fact in
our $l_{2}=0$ case the Hubble parameter, the deceleration and EoS
parameters and the scalar field simplify to
\begin{eqnarray}
& & H = \frac{n+1}{t}, \label{Hubble-all} \\
& & q= -\frac{n}{n+1}, \label{dec-3} \\
& & w=-\frac{3 n + 1}{3\left(n + 1\right)}, \label{case-2-eos} \\
& & \Phi(t) =\frac{6 n^2 }{\lambda} t^{-2} \label{sclr-fld4}
\end{eqnarray}
and we can confirm that $q$ is proportional to $1+3w$; in
particular, for $n=-1/2$ Eq.(\ref{case-2-eos}) gives $w=1/3$ and
for $n=-1/3$, it gives $w=0$. For $n \rightarrow 0$, the
deceleration parameter vanishes. We know that empty universes can
have this property, and we see that the scalar field vanishes in
this case. For $n > 0$, the effective fluid is equivalent to
quintessence/quiessence ($-1<w<-1/3$), approaching cosmological
constant ($w=-1$) as $n \rightarrow \infty$. Only for $n<-1$ do we
have phantom behavior ($w<-1$), and this is the case where a Big
Rip appears (for negative times$^1$, at least).

The $n=-1$ case is particularly interesting. For this case, $q$
and $w$ seem to diverge, while in Table \ref{tb:l2zero} this
subcase does not seem to be more problematic than others. But, as
$n$ approaches $-1$, the scale factor $a(t)$ approaches a constant
function, therefore $\dot{a}(t)$ and $\ddot{a}(t)$ approach zero.
Hence the deceleration parameter, which has division by
$\dot{a}^{2}$ in its definition, is not a good parameter to use in
this limit. Also, since this universe corresponds to Minkowski
space, the stress-energy-momentum tensor should vanish. Let us
calculate the energy density and isotropic pressure, using
Eq.(\ref{sclr-fld4}) and the potential expression in
Eq.(\ref{pot}), in expressions (\ref{en-dens}) and (\ref{pres}):
\begin{eqnarray}
& & \rho^{\Phi} = \frac{18}{\lambda} n^2 (n+1)^2 \, t^{-4},  \label{ed-l2} \\
& & p^{\Phi} = -\frac{18}{\lambda} n^2 (n+1) (n+\frac{1}{3}) \, t^{-4}.  \label{p-l2}
\end{eqnarray}
These expressions verify that $n=0$ corresponds to empty
universes. It can also be seen that {\it both} the density and
pressure vanish for $n=-1$, although the field does not vanish! Of
course, this also makes $w$ meaningless in this subcase.

\subsubsection{The $l_{1}=0$ case.} \label{powerl1zero}

Since $l_1= \lambda /(6 l_0^2)$, in this case the potential
$U(\Phi)=\lambda \Phi^2$ vanishes, therefore this case is
equivalent considering the scalar field to be massless. Now $A(t)
\propto t^{1-2n}$, $B(t) \propto t^{1-n/2}$ and $a(t) \propto
t^{1-n}$, so that the critical $n$ values are 1/2, 1 and 2; and we
are led to Table \ref{tb:l1zero} listing the subcases.

\begin{table}[htbp!]
\centering
\begin{tabular}{|c|lcccr|}
\hline
\backslashbox{$n$}{$t$}  & $-\infty$ &   & $0$ &  & $+\infty$ \\ \hline

$n < 1/2$     & [$\infty,\infty,\infty$]I    & & BC[0, 0, 0]BB     & & I[$\infty,\infty,\infty$] \\ \hline

$n = 1/2$     & [C, $\infty, \infty $]2I & & 2BC[C, 0, 0]2BB & & 2I[C,$\infty, \infty$] \\ \hline

$1/2 < n < 1$ & [0, $\infty$, $\infty$]IP$_{\rm a}$ & & cBC[$\infty$, 0, 0]cBB
 & & IP$_{\rm a}$[0, $\infty$, $\infty$] \\ \hline

$n = 1$       & [0, $\infty$, C]IP$_{\rm b}$        & & BD[$\infty$, 0, C]iBD
 & & IP$_{\rm b}$[0, $\infty$, C] \\ \hline

$1 < n < 2$   & [0, $\infty$, 0]IP$_{\rm c}$  &  & cBR[$\infty$, 0, $\infty$]ciBR
 &  & IP$_{\rm c}$[0, $\infty$, 0] \\ \hline

$n = 2$   & [0, C, 0]1Z    & & 1BR[$\infty$, C, $\infty$]1iBR & & 1Z[0, C, 0] \\ \hline

$n > 2$   & [0, 0, 0]Z    & & BR[$\infty, \infty, \infty$]iBR & & Z[0, 0, 0] \\ \hline

\end{tabular}
\caption{Universes derived for  $l_{1} = 0$, where $\nu(t)$ is a
power of $t$. Each line shows a timeline from $-\infty$ to
$\infty$, but may represent {\em multiple} universes delimited by
singularities, shown by square brackets. The expressions in the
brackets refer to $A(t)$, $B(t)$ and $a(t)$. Also, I: infinite
scale factors at infinite future or past, BR: Big Rip, i: Inverse,
BC: Big Crunch, BB: Big Bang, 1 or 2: only one or two of the
spatial dimensions, c: cigar-type, BD: Big Draw, IP$_{\rm a}$ :
"Infinite pancake" of type a, etc., Z: vanishing scale factors at
infinite future or past [Also see text and the note about negative
time on page \pageref{negNrest}]. \label{tb:l1zero}}
\end{table}

\noindent Let us note that

\begin{itemize}

\item For $n = 1/2$, we get universes evolving in the $y$ and $z$
dimensions only, either from infinite scale factors to a BC, or
from a BB to infinite scale factors, in infinite time, possibly to
be called 2I-2BC and 2BB-2I for two-dimensional

\item For $1/2<n<1$, $A(t)$ diverges at $t=0$, while the other
scale factors vanish. This singularity is extremely anisotropic:
As the universe contracts towards a BC in the $y$ and $z$
directions, it actually expands infinitely in the $x$ direction!
But this expansion is not enough to prevent the vanishing of a
comoving volume, and in this sense this event is a BB/BC.
Singularities where one dimension diverges while two shrink to
zero are called "cigar-type", so we might call this event a
"cBC/cBB".

Again for $1/2<n<1$, $A(t)$ vanishes as $t\rightarrow \pm \infty$, while the other scale factors do not. This type of singularity is also extremely anisotropic, and is called a "pancake" singularity. In this particular subcase, the other scale factors diverge, so the singularity is called an {\em infinite pancake}. This type of singularity appears in the next two subcases too, but the behavior of the average scale factor $a(t)$ is diferent in each subcase. So we classified the "infinite pancakes" accordingly. (In \cite{roynarsin}, a classification is made according to the behavior of the three separate scale factors of a Bianchi I spacetime.)

\item  For $n=1$, $A(t)$ diverges, $a(t)$ goes to a constant,
$B(t)$ vanishes at $t=0$. Although this $t=0$ event is also a
cigar-type singularity, since the comoving volume neither vanishes
nor diverges, it is not a BB/BC or BR/iBR. We suggest the names
Big Draw (BD)\footnote{since the volume of a metal drawn to
produce a wire does not change.} and Inverse Big Draw (iBD).

\item For $n = 2$, the universe evolves in the $x$ dimension only,
either from vanishing scale factor to a BR, or from an iBR to
vanishing scale factor, in infinite time. For $n
> 2$, this occurs in all three dimensions.

\end{itemize}

For $l_{1}=0$, the mean Hubble parameter, the deceleration parameter, the EoS parameter and the scalar field simplify respectively to
\begin{eqnarray}
& & H = \frac{1-n}{t},  \label{meanHubble3-l1} \\
& & q = \frac{n}{1-n},  \label{dec-2-l1} \\
& &  w= \frac{-5n +2}{3(n -2)},  \label{case-2-2-eos-0}  \\
& & \Phi(t) = \frac{ a_1^3 n^2}{l_0^2 l_2}
t^{3n-2}. \label{sclr-fld5}
\end{eqnarray}

There seem to be problems for $n=1$ and $n=2$, while in Table
\ref{tb:l1zero} these cases do not seem to be more problematic
than others. The apparent divergence of $q$ as $n=1$ is understood
as for the $n=-1$ line of Table \ref{tb:l2zero}: $a(t)$ becomes
constant. As discussed at the end of Section \ref{eos}, the $w$
parameter is not very useful when there is anisotropy\footnote{As
a specific example, consider $n=7/5$, which gives $w=25/9$
according to (\ref{case-2-2-eos-0}), which should give
deceleration, but gives $q=-7/2$, that is, acceleration according
to (\ref{dec-2-l1}).}, but one might still ask why it diverges for
$n=2$. Let us again calculate the energy density and isotropic
pressure of the effective fluid for our subcase, recalling that
the potential vanishes. Then, using Eq.(\ref{sclr-fld5}), we get
\begin{eqnarray}
& & \rho^{\Phi} = \frac{3 a_1^3}{4 l_0^2 l_2} n^2 (n-2) (3n -2) \, t^{3n -4},  \label{ed-l1} \\
& & p^{\Phi} = \frac{a_1^3}{4 l_0^2 l_2} n^2 (2-5n) (3n -2) \, t^{3n -4}.  \label{p-l1}
\end{eqnarray}
explaining the divergence of $w$ by the vanishing of $\rho^\Phi$
for $n=2$. We also see that the density and pressure {\em both}
vanish (other than for the trivial, isotropic $n=0$ case) for
$n=2/3$, due to the constancy of the field and vanishing of the
potential.

\subsubsection{The case with both $l_{1}$ and $l_{2}$ nonzero.}

For this subcase, the limiting behaviors of the scale factors
mentioned on page~\pageref{powerScaleLimits} show that they
diverge near infinities for all $n$. Moreover, at late time, the
universe expands with acceleration, again for all $n$, as can be
seen from the asymptotic behavior mentioned after
Eq.(\ref{meanHubble3}). But the behavior of the scale factors near
zero determines the critical values of $n$ as -1, 1/2, 1 and 2,
leading to the subcases shown  in Table \ref{tb:l1l2nonzero}.
Outstanding features are

\begin{table}[htbp!]
\centering
\begin{tabular}{|c|l|lcccccccr|}
\hline
\multicolumn{2}{|r|}{\backslashbox{$n, t_{1}$}{$t$}}  & $-\infty$ &
 & $-|t_{1}|$ &   & $0$  &  & $|t_{1}|$ & & $+\infty$ \\ \hline

$n < -1$ & No $t_1$ & [$\infty$]I &  & \multicolumn{2}{c}{---B---} &
BR[$\infty$]iBR & \multicolumn{2}{c}{---B---} & & I[$\infty$] \\ \cline{2-11}

         & $t_{1}<0$ & [$\infty$]I & & BC[0]BB &  &
BR[$\infty$]iBR & \multicolumn{2}{c}{---B---} & & I[$\infty$] \\ \cline{2-11}

         & $t_{1}>0$ & [$\infty$]I & & \multicolumn{2}{c}{---B---} &
BR[$\infty$]iBR &  & BC[0]BB & & I[$\infty$] \\ \hline

$n = -1$ & No $t_1$ & [$\infty$]I &            & & &
(C) & & \multicolumn{2}{l}{---B---$^{*}$}  & I[$\infty$] \\ \cline{2-11}

         & $\exists \; t_{1}$ & [$\infty$]I & & &  &
(C) &  & BC[0]$^{*}$BB & & I[$\infty$] \\ \hline

$-1< n<0$ & No $t_1$ & [$\infty$]I &            & & &
BC[0]BB & & & & I[$\infty$] \\ \cline{2-11}

         & $t_{1}<0$ & [$\infty$]I & & BC[0]BB &  &
BC[0]BB & & & & I[$\infty$] \\ \cline{2-11}

         & $t_{1}>0$ & [$\infty$]I & & & &
BC[0]BB &  & BC[0]BB & & I[$\infty$] \\ \hline

\multicolumn{2}{|c|}{$n \rightarrow 0$} & [$\infty$]I & \multicolumn{3}{c}{linear  contraction} &
BC[0]BB & \multicolumn{3}{c}{linear  expansion} & I[$\infty$] \\ \hline

$0<n<1/2$ & No $t_1$ & [$\infty$]I &            & & &
BC[0]BB & & & & I[$\infty$] \\ \cline{2-11}

         & $t_{1}<0$ & [$\infty$]I & & BC[0]BB &  &
BC[0]BB & & & & I[$\infty$] \\ \cline{2-11}

         & $t_{1}>0$ & [$\infty$]I & & & &
BC[0]BB &  & BC[0]BB & & I[$\infty$] \\ \hline

$n=1/2$ & No $t_1$ & [$\infty$]I &            & & &
2BC[C,0]2BB & & & & I[$\infty$] \\ \cline{2-11}

         & $t_{1}<0$ & [$\infty$]I & & BC[0]BB &  &
2BC[C,0]2BB & & & & I[$\infty$] \\ \cline{2-11}

         & $t_{1}>0$ & [$\infty$]I & & & &
2BC[C,0]2BB &  & BC[0]BB & & I[$\infty$] \\ \hline

$1/2<n<1$ & No $t_1$ & [$\infty$]I &            & & &
cBC[$\infty$,0,0]cBB & & & & I[$\infty$] \\ \cline{2-11}

         & $t_{1}<0$ & [$\infty$]I & & BC[0]BB &  &
cBC[$\infty$,0,0]cBB & & & & I[$\infty$] \\ \cline{2-11}

         & $t_{1}>0$ & [$\infty$]I & & & &
cBC[$\infty$,0,0]cBB &  & BC[0]BB & & I[$\infty$] \\ \hline

$n=1$ & No $t_1$ & [$\infty$]I &            & & &
BD[$\infty$,0,C]iBD & & & & I[$\infty$] \\ \cline{2-11}

         & $t_{1}<0$ & [$\infty$]I & & BC[0]BB &  &
BD[$\infty$,0,C]iBD & & & & I[$\infty$] \\ \cline{2-11}

         & $t_{1}>0$ & [$\infty$]I & & & &
BD[$\infty$,0,C]iBD &  & BC[0]BB & & I[$\infty$] \\ \hline

$1<n<2$ & No $t_1$ & [$\infty$]I &            & \multicolumn{2}{c}{---B---} &
cBR[$\infty$,0,$\infty$]ciBR & \multicolumn{2}{c}{---B---} & & I[$\infty$] \\ \cline{2-11}

         & $t_{1}<0$ & [$\infty$]I & & BC[0]BB &  &
cBR[$\infty$,0,$\infty$]ciBR & \multicolumn{2}{c}{---B---} & & I[$\infty$] \\ \cline{2-11}

         & $t_{1}>0$ & [$\infty$]I & & \multicolumn{2}{c}{---B---} &
cBR[$\infty$,0,$\infty$]ciBR &  & BC[0]BB & & I[$\infty$] \\ \hline

$n=2$ & No $t_1$ & [$\infty$]I &            & \multicolumn{2}{c}{---B---} &
1BR[$\infty$,C,$\infty$]1iBR & \multicolumn{2}{c}{---B---} & & I[$\infty$] \\ \cline{2-11}

         & $t_{1}<0$ & [$\infty$]I & & BC[0]BB &  &
1BR[$\infty$,C,$\infty$]1iBR & \multicolumn{2}{c}{---B---} & & I[$\infty$] \\ \cline{2-11}

         & $t_{1}>0$ & [$\infty$]I & & \multicolumn{2}{c}{---B---} &
1BR[$\infty$,c,$\infty$]1iBR &  & BC[0]BB & & I[$\infty$] \\ \hline

$n > 2$ & No $t_1$ & [$\infty$]I &            & \multicolumn{2}{c}{---B---} &
BR[$\infty$]iBR & \multicolumn{2}{c}{---B---} & & I[$\infty$] \\ \cline{2-11}

         & $t_{1}<0$ & [$\infty$]I & & BC[0]BB &  &
BR[$\infty$]iBR & \multicolumn{2}{c}{---B---} & & I[$\infty$] \\ \cline{2-11}

         & $t_{1}>0$ & [$\infty$]I & & \multicolumn{2}{c}{---B---} &
BR[$\infty$]iBR &  & BC[0]BB & & I[$\infty$] \\ \hline

\end{tabular}
\caption{Universes derived for  $l_{1} \neq 0$ and $l_{2} \neq 0$,
where $\nu(t)$ is a power of $t$. Each line shows a timeline from
$-\infty$ to $\infty$, but may represent {\em multiple} universes
delimited by singularities, shown by square brackets. A single
expression in the brackets refers to all scale factors, two
expressions refer to $A(t)$ and $B(t)$, three expressions refer to
$A(t)$, $B(t)$ and $a(t)$. Also, I: infinite scale factors at
infinite future or past, B: Bounce, BR: Big Rip, i: Inverse, BC:
Big Crunch, BB: Big Bang, 1 or 2: only one or two of the spatial
dimensions, c: cigar-type, BD: Big Draw [See text, and also note
about negative time on page \pageref{negNrest}. $^{*}$Sign of
bounce time or $t_{1}$ may be positive {\em or} negative].
\label{tb:l1l2nonzero}}
\end{table}

\begin{itemize}

\item The comoving volume diverges at $t=0$ for $n<-1$ and for
$n>1$. Since it also diverges as $t \rightarrow \pm \infty$, this
means that it must go through a minimum between two divergences
--- a Bounce (B). If $t_{1}$ does not exist, there will be a
Bounce each in the $t<0$ universe$^1$ and the $t>0$ universe;
whereas if $t_{1}$ does exist, it will essentially bring one of
the minima down to zero, inserting a BC/BB, effectively splitting
one of the universes into two. Hence the first line of Table
\ref{tb:l1l2nonzero} has an Infinity-Bounce-Big Rip (I-B-BR)
universe and  an iBR-B-I universe; the second line an I-BC
universe, a finite-lifetime BB-BC universe and an iBR-B-I
universe; and so on.

\item  The third line features an iBR-BC universe, i.e. a
finite-lifetime universe which starts with infinitely large scale
factors that immediately contract to finite values, and keep on
contracting to a BC.

\item  For $n=-1$, the scale factors become constant at $t=0$.
Therefore the subcase without $t_{1}$ features the only universe
of the table without beginning$^1$ or end, a bouncing universe.

\item The $n \rightarrow 0$ case is the same as the corresponding
case of Table \ref{tb:l2zero}. In fact, it is isotropic.

\item For all cases with $n>0$, the singularity at $t=0$ is
qualitatively the same as in the corresponding $n$ value Table
\ref{tb:l1zero}.

\end{itemize}

Interestingly, a wide range of cosmological possibilities,
including late-time acceleration, can be reproduced by using a
power of $t$ as the seed function. This even includes the
isotropic case, where the time-dependence of the source scalar
field does not change with $n$: $\Phi(t) \propto t^{-2}$. This
surprising richness, including the possibility of vanishing
energy-momentum tensor while the field is nonzero, seems to be due
to the coupling coefficient selected by the Hessian determinant
condition, and the potential selected by the Noether symmetry
approach.

\subsection{Exponential seed}
\label{ustel}

When we use $\nu(t)=a_1 e^{kt}$ where $a_2$ and $k$ are non-zero
constants, in Eqs.(\ref{metric-A1})-(\ref{sclr-fld}),
(\ref{scale-f1}), (\ref{meanHubble-genel}) and (\ref{decparam});
the scale factors, the scalar field, the mean Hubble parameter and
the deceleration parameter become
\begin{eqnarray}
& & A(t) = \frac{l_0}{k a_2^2} \left( l_1 a_2^3 e^{3 kt} +
l_2 \right) e^{-2kt}, \label{scale-fX}\\
& & B(t) = \frac{l_0 }{k \sqrt{a_2}} \left( l_1 a_2^3 e^{3 kt} + l_2 \right)^{1/2} e^{-kt/2}, \label{scale-fYZ}\\
& & a(t) = \frac{l_0 }{k a_2} \left( l_1 a_2^3 e^{3 kt} +
l_2 \right)^{2/3} e^{-kt}, \label{scale-f2}\\
& & \Phi(t) = \frac{ k^2 a_2^3}{l_0^2} \frac{ e^{3kt}}{l_1 a_2^3
e^{3kt} + l_2}, \label{sclr-fld2}\\
& &  H= k \left[ \frac{l_1 a_2^3 e^{3kt} - l_2}{ l_1 a_2^3 e^{3kt} + l_2 } \right], \label{meanHubble2}\\
& &  q = - \frac{l_1^{2} a_2^6
e^{6k t} + 4 l_1 l_2 a_1^3  e^{3 kt}+l_{2}^{2}}{\left(l_1 a_1^3
e^{3k t} - l_2\right)^2 }. \label{dec-1}
\end{eqnarray}

Unlike in Section \ref{powerlaw}, these scale factors are never
singular at $t=0$, however the $(l_1 a_2^3 e^{3kt} + l_2)$ terms
in the scale factors can vanish at some finite $t$ value $t_{2}$,
if $l_{1}$ and $l_{2}$ are both nonzero. Since they do not diverge
at finite time, there is no Big Rip.

In that case, the scale factors diverge exponentially at both
infinities, regardless of the signs of $l_{1}$, $l_{2}$ and $k$.
Therefore, if $t_{2}$ does not exist, the universe is a bouncing
universe, infinite in both time directions. This solution is
qualitatively similar to the one in the fourth line of Table
\ref{tb:l1l2nonzero}. If $t_{2}$ does exist, then the solution
represents two universes; one collapsing from infinite scale
factors to a Big Crunch, one expanding from a Big Bang to infinite
scale factors. This solution is qualitatively similar to the one
in the fifth line of Table \ref{tb:l1l2nonzero}. For all solutions
(with nonzero $l_{1}$ and $l_{2}$), at late time, $H \rightarrow
|k|$ and $q \rightarrow -1$, confirming the de Sitter-like
asymptotic behavior of the universe(s). For positive $k$, the
scalar field $\Phi$ approaches a constant.

If either one of $l_{1}$ and $l_{2}$ is zero, the universe either
expands or contracts exponentially, depending on the sign of $k$
and on which constant vanishes. In these cases, $q = -1$ at all
times, and for $l_{2}=0$ (isotropic), $\Phi$ is constant. A
constant scalar field is the standard (simplest) way in the
literature for causing exponential expansion, and in the isotropic
case can be interpreted as creating a cosmological constant. For
positive $k$, the solution approaches isotropy at late times.

According to recent astrophysical data, the deceleration parameter
of the universe lies in the interval $-1.72 < q < -0.58$
\cite{Melchiori}. In all models of this subsection at late time
$q$ converges to $-1$, so in this sense these models are
consistent with the observational results.

\subsection{Shifted-reversed power seed}
\label{BigRipSeed}

We may also use $\nu(t)=\frac{a_{3}}{(t_{c}-t)^{m}}$, where $a_3$,
$t_{c}$ and $m$ are non-zero constants, since this function is
similar to the FRW Big Rip models. Then
Eqs.(\ref{metric-A1})-(\ref{sclr-fld}) become
\begin{eqnarray}
& & A(t) = \frac{l_0}{a_{3}^{2} m} (t_{c}-t)^{1+2m} \left[l_1
a_3^3 (t_{c}-t)^{-3m}+l_2\right], \label{rip-A} \\ & & B(t)=
\frac{-l_0}{m \sqrt{a_{3}}} (t_{c}-t)^{1+m/2}  \left[l_1 a_3^3
(t_{c}-t)^{-3m}+l_2\right]^{1/2},  \label{rip-B}
\end{eqnarray}
\begin{eqnarray}
& & \Phi(t) = \frac{ a_3^3 m^2 (t_{c}-t)^{-3m-2}}{l_0^2 \left[l_1
a_3^3 (t_{c}-t)^{-3m}+l_2 \right]}. \label{rip-sclrfld}
\end{eqnarray}
The substitutions $m \rightarrow -n$ and $t \rightarrow t_{c}-t$
bring these expressions into the same form as
Eqs.(\ref{powerlaw-A-B}) and (\ref{sclr-fld3}). Therefore all
solutions from Section \ref{powerlaw} apply, we just have to read
the lines of the tables in that section from right to left, and
switch the sign of $n$ (and of course, shift any "i" designation
to the other side of the singularity). So this seed does not only
give Big Rip models (see Ref.\cite{noj005} for detailed
discussion).

\subsection{Other seeds and parameter choices}
\label{OtherSeeds}

Of course, $\nu(t)$ being a free function, an
infinite number of choices is possible, not to mention the freedom
in the choices of $l_{1}$ and $l_{2}$.  Among these, we are
naturally interested in those choices that give simple metric
functions $A(t)$ and $B(t)$.

One way of achieving this would be to set $l_{1}$
or $l_{2}$ equal to zero. But $l_{2}=0$ takes us to isotropic
solutions, which are not in line with the emphasis of the present
work. The choice $l_{1}=0$, although it will give simple-looking
solutions, limits us to the case of massless field, and there
seems to be no motivation for choosing a particular solution above
others, except possibly $\nu(t)=t$ (since the comoving volume
stays constant), but this is already covered in subsection
\ref{powerl1zero}.

One can also choose combinations to make both $\nu(t)$ and $(l_1
\nu^3 + l_2)$ simple. For example, choosing
$\nu(t)=\cos^{2/3}(t)$, $l_{0}=-2/3$, $l_{1}=-1$ and $l_{2}=1$
yields
\begin{equation} \label{metric-trig}
ds^2 = -dt^2 + \cos^{4/3} (t) \left[ \tan^2 (t) \, dx^2 + dy^2 +
dz^2 \right]
\end{equation}
a metric with scale factors
\begin{eqnarray} 
& &  A(t) =  \frac{\sin (t)}{\cos^{1/3}(t)}  \\
& &  B(t) =  \cos^{2/3} (t)  \\
& &  a(t) =  \left[\sin (t) \cos(t)\right]^{1/3}
\end{eqnarray}
 i.e. a set of universes that start with a 1-D
Big Bang and end with a cigar-like Big Crunch (for $0 \leq t \leq
\pi/2$, in appropriate units of time) or start with a cigar-like
Big Bang and end with a 1-D Big Crunch (for $\pi/2 \leq t \leq
\pi$).

Another solution,
\begin{equation} \label{metric-hyp}
 ds^2 = -dt^2 + \sinh^{4/3} t \left( \coth^2 t \, dx^2 + dy^2 + dz^2 \right)
\end{equation}
is mathematically similar, but describes
universes with infinite lifetime, for example the $t>0$ universe
starts with a cigar-like Big Bang and approaches de Sitter-like
isotropically expanding universe at late time.

\section{Solutions in Einstein frame}
\label{einstein}

It is known that a scalar-tensor theory can be transformed to the
so-called {\em Einstein frame}, where the gravitational scalar
field becomes minimally coupled to curvature. The price to pay for
this simplification is the equivalence principle: Massive point
particles do not follow geodesics any more, in contrast to the
{\em Jordan frame}, which we used so far in this
work~\cite{fm,faraoni}. For the BD theory, the case $W=-3/2$ is
special: We derived it from the vanishing of the Hessian
(\ref{hessian}), it represents a fixed-point of the conformal
transformation~\cite{faraoni}, and marks the boundary beyond which
ghosts appear~\cite{dabrowski}. In the Einstein frame, this
specialness is reflected in the gravitational scalar field
becoming non-dynamical. Hence, it would be interesting to look at
our solutions also in the Einstein frame.

The conformal transformation
\begin{eqnarray}
& & g_{ab} =\Omega^{-2} \tilde{g}_{ab}, \quad g^{ab} =\Omega^{2}
\tilde{g}^{ab},  \label{conftr}
\end{eqnarray}
takes us from the Jordan frame to the Einstein frame. Defining
$\Phi = e^{-\sigma}$, the $W=-3/2$ BD action given in Eq.
(\ref{action-bd}) takes the form
\begin{eqnarray}
\mathcal{A_{BD}} & = & \int{d^4 x \sqrt{-g} e^{-\sigma} \left[ R+
\frac{3}{2}\sigma_{c} \sigma^{c} - U_1(\sigma)) \right]}
\label{action-bd-2}
\end{eqnarray}
where $U_1 (\sigma)= U(\Phi)/\Phi$. It is easy to show that under a choice of
a conformal factor
\begin{eqnarray}
& & \Omega = e^{-\sigma/2},  \label{omega}
\end{eqnarray}
the above action (\ref{action-bd-2}) transforms
into
\begin{eqnarray}
\mathcal{A_{E}} & = & \int{d^4 x \sqrt{-\tilde{g}} \left[
\tilde{R} - U_2(\sigma)) \right]}. \label{action-ein}
\end{eqnarray}
This is exactly the Einstein-Hilbert action with a potential $U_2
(\sigma)= e^{\sigma} U_1 (\sigma)= U(\Phi)/\Phi^2 $, and using the
potential obtained in (\ref{pot}) it takes the form $U_2 (\sigma)
= \lambda$, where $\lambda$ is a constant and could be interpreted
as the cosmological constant in Einstein frame. In this case, the
(transformed) scalar field $\sigma$ does not appear in the action,
hence the scalar field is non-dynamical as stated in the beginning
of this section; and the field equations take the form
\begin{eqnarray}
& & \tilde{R}_{ab} = \frac{\lambda}{2} \tilde{g}_{ab}.
\label{f-eq-ein}
\end{eqnarray}
where the Ricci tensor $\tilde{R}_{ab}$ refers to the transformed metric $\tilde{g}_{ab}$.

The LRS Bianchi I spacetime can be brought back to its original form
\begin{equation} \label{metric-b1tr}
ds^2 = -d\tilde{t}^2 + \tilde{A}^2 dx^2 + \tilde{B}^2 \left( dy^2
+ dz^2 \right),
\end{equation}
by a simple coordinate transformation after the conformal
transformation.  The transformations of time coordinate and scale
factors from the Jordan frame to the Einstein frame are given by
\begin{eqnarray}
& & \tilde{t} = \int{\sqrt{\Phi} dt}, \quad \tilde{A} =
\sqrt{\Phi} A, \quad \tilde{B} = \sqrt{\Phi} B.
\label{tr-metric-coeff}
\end{eqnarray}
For this spacetime in Einstein frame the field equations (\ref{f-eq-ein}) give
\begin{eqnarray}
& & 2 \frac{\tilde{A}' \tilde{B}'}{\tilde{A} \tilde{B}} + \left(
\frac{\tilde{B}'}{\tilde{B}} \right)^2  = \frac{\lambda}{2},
\label{f-eq-ein-1}  \\& & 2 \frac{\tilde{B}''}{\tilde{B}} + \left(
\frac{\tilde{B}'}{\tilde{B}} \right)^2  = \frac{\lambda}{2},
\label{f-eq-ein-2} \\ & & \frac{\tilde{A}''}{\tilde{A}} +
\frac{\tilde{B}''}{\tilde{B}} + \frac{\tilde{A}'
\tilde{B}'}{\tilde{A} \tilde{B}} = \frac{\lambda}{2},
\label{f-eq-ein-3}
\end{eqnarray}
where the prime represents derivative with respect to tilted time
coordinate $\tilde{t}$. These equations can be solved exactly, giving
\begin{eqnarray}
& & \tilde{A} = c_3 \left( \tilde{t} + c_1 \right)^{-1/3}, \quad
\tilde{B} = c_2 \left( \tilde{t} + c_1 \right)^{2/3},
\label{tr-metric-coeff-t-0}
\end{eqnarray}
for $\lambda =0$;
\begin{eqnarray}
& & \tilde{A} = c_3 \sinh (k\tilde{t} +c_1) \cosh^{-1/3}
(k\tilde{t} +c_1), \nonumber \\
& &  \tilde{B} = c_2 \cosh^{2/3}(k\tilde{t} +c_1),
\label{tr-metric-coeff-t-1}
\end{eqnarray}
for $\lambda = 8 k^2 / 3 > 0$;
\begin{eqnarray}
& & \tilde{A} = c_3 \sin (k\tilde{t} +c_1) \cos^{-1/3}
(k\tilde{t} +c_1), \nonumber \\
& & \tilde{B} = c_2 \cos^{2/3}(k\tilde{t} +c_1),
\label{tr-metric-coeff-t-2}
\end{eqnarray}
for $\lambda = -8 k^2 / 3 < 0$. We note here  that the obtained
scale factors given in (\ref{tr-metric-coeff-t-0}) represent the
well known Kasner solution. We would like to check the consistency
between these solutions and those in the Jordan frame,
(\ref{metric-A1})-(\ref{sclr-fld}). For given $\lambda$, the
Jordan frame solutions contain the arbitrary function $\nu(t)$ and
two arbitrary constants $l_{0}$ and $l_{2}$ ($l_{1}$ is not
independent). The Einstein frame solutions contain one arbitrary
function $\Phi(\tilde{t})$ (which does not appear in the metric,
however) and three arbitrary constants, $c_{1}$, $c_{2}$ and
$c_{3}$. To show consistency, we need to transform the Jordan
frame solutions to the Einstein frame and show that they agree
with solution found in that frame.

Applying the transformations in (\ref{tr-metric-coeff}) to
(\ref{metric-A1}) and (\ref{metric-B1}), using (\ref{sclr-fld}),
we get
\begin{eqnarray}
& & \tilde{A} = \sqrt{\frac{l_{1} \nu(t){^3} + l_{2}}{\nu(t)}}, \label{a_trans} \\
& & \tilde{B} = \nu(t) \label{b_trans}
\end{eqnarray}
Using (\ref{b_trans}) in (\ref{sclr-fld}), we can write for $\Phi$
in the transformed coordinates
\begin{equation}
\Phi(t) = \frac{\tilde{B} \dot{\tilde{B}}^{2}}{l_0^{2} \left(l_{1}
\tilde{B}{^3} + l_{2}\right)}, \label{phi_trans}
\end{equation}
but using (\ref{tr-metric-coeff}),
\begin{equation}
\dot{\tilde{B}} = \frac{d\tilde{B}}{dt} =
\frac{d\tilde{B}}{d\tilde{t}}\frac{d\tilde{t}}{dt} = \tilde{B}'
\sqrt{\Phi} \Longrightarrow \Phi = \frac{\tilde{B} \tilde{B}'^{2}
\Phi}{l_0^{2} \left(l_{1} \tilde{B}{^3} + l_{2}\right)},
\label{phi_trans}
\end{equation}
so $\Phi$ disappears from the equation: it cannot be determined in
the transformed coordinates. This was to be expected, since the
scalar field is absent from the action (\ref{action-ein}). The
arbitrariness (or information) in $\nu(t)$ in the Jordan frame has
shifted to $\Phi(\tilde{t})$ in the Einstein frame.

Returning to  (\ref{phi_trans}) and using the solution found for
$\tilde{B}(\tilde{t})$ , e.g. for positive $\lambda$,
\begin{equation}
l_0^{2} \left(l_{1} c_{2}^{3} \cosh^{2}
(k\tilde{t} +c_1) + l_{2}\right) = c_{2}^{3} \frac{4}{9}k^2 \sinh^{2}
(k\tilde{t} +c_1) ,
\end{equation}
where it should be noted that  $k^2 = 3 \lambda / 8 = 9 l_{1}
l_{0}^{2}/4$ (see after eq.(\ref{sol1})). This will hold, if
$c_{2}$ is chosen such that $l_{2}$ is equal to $-l_{1}
c_{2}^{3}$. This choice can be made, since obviously the set $\{
c_{1}, c_{2}, c_{3}\}$ cannot be independent of the set $\{ l_{0},
l_{2}\}$: If two solutions describe the same physical reality,
the parameters (arbitrary constants) of one  should be expressible
in terms of the parameters of the other, although information
could hide in the arbitrary function(s) in this case.

Similarly, (\ref{a_trans}) will agree with
(\ref{tr-metric-coeff-t-1}), if $c_{3}$ is properly chosen; but
$c_{1}$ is not related to $l_{0}$ or $l_{2}$, it comes from the
integration in the first term in (\ref{tr-metric-coeff}). The same
calculations can be made for the cases of negative and vanishing
$\lambda$, establishing consistency of Einstein frame solutions
with the Jordan frame solutions.

For example, we can transform the solutions of subsection 6.1.1
($l_2 = 0$, isotropic) as
\begin{equation}
\sqrt{\Phi}= \frac{3 n}{2 k t} \Longrightarrow t = t_{1}
e^{\frac{2 k}{3 n}\tilde{t}}, \;\; \tilde{A}= \frac{2 k a_{1}}{3
l_{0}} t_{1}^{n} e^{\frac{2 k}{3}\tilde{t}}, \;\; \tilde{B} =
a_{1} t_{1}^{n} e^{\frac{2 k}{3}\tilde{t}}
\label{tr-metric-coeff-1}
\end{equation}
and those of subsection 6.1.2 ($l_{1}=0$) as
\begin{eqnarray}
& & \sqrt{\Phi}= \left[\frac{n^{2} a_{1}^{3}}{l_{0}^{2}
l_{2}}\right]^{1/2} t^{(3/2)n-1} \Longrightarrow \nonumber \\
& & t = t_{0} (\tilde{t}+\tilde{c_{1}})^{2/(3n)}, \;\; \tilde{A} =
A_{0} (\tilde{t}+\tilde{c_{1}})^{-1/3}, \;\; \tilde{B} = B_{0}
 (\tilde{t}+\tilde{c_{1}})^{2/3},
\label{tr-metric-coeff-2}
\end{eqnarray}
where
\begin{equation}
t_{0} = \left[\frac{9 l_{0}^{2} l_{2}}{4
a_{1}^{3}}\right]^{1/(3n)}, \;\; A_0=\left[\frac{2
l_{2}}{3l_{0}}\right]^{1/3} \;\;{\rm and} \;\; B_0=\left[\frac{9
l_{0}^2 l_{2}}{4}\right]^{1/3}.
\end{equation}
Both solutions (\ref{tr-metric-coeff-1}) and
(\ref{tr-metric-coeff-2}) identically satisfy the Einstein frame
equations (\ref{f-eq-ein-1})-(\ref{f-eq-ein-3}), if one recalls
that $l_{1}=0$ implies vanishing of $\lambda$.The constants
$t_{1}$ in (\ref{tr-metric-coeff-1}) and $\tilde{c_{1}}$ in
(\ref{tr-metric-coeff-2}) correspond to the constants $c_{1}$ that
appear in (\ref{tr-metric-coeff-t-0})-(\ref{tr-metric-coeff-t-2}).

To summarize, the solutions we found in the Einstein frame,
although containing only three parameters, are the transformed
versions of the solutions in the Jordan frame, which contained an
arbitrary function. This correspondence between a solution for the
metric in the Einstein frame and an infinite number of metrics in
the Jordan frame is possible, since the scalar field, which
determines the transformation between the two frames, is arbitrary
in the Einstein frame.

\section{Concluding remarks}
\label{conc}

In this paper we have examined the scalar-tensor Brans-Dicke
theory of gravity for LRS Bianchi type I spacetime admitting
Noether symmetry. This symmetry approach is important because it
provides us with a theoretical motivation to select a region of
the solution space (For other motivations, leading to different
regions of the solution space for the Bianchi type I spacetimes,
see e.g. Refs.\cite{rodrigues}-\cite{calogero2}). The Lagrangian
density (\ref{lagra}) of LRS Bianchi type I becomes degenerate for
$W(\Phi)=-3/2$, when the Brans-Dicke coupling function $F(\Phi) =
\Phi$ is used. This degeneracy is required for nontrivial
solutions, hence we use this value as the BD parameter. The
existence of Noether symmetry also restricts the form of potential
$U(\Phi)$, and allows us to find a transformation given by
(\ref{i-trns}) in which the metric potentials and the scalar field
are stated in terms of new dynamical variables ($\mu, \nu, u$),
where the variable $\mu$ is cyclic. Under the transformation
(\ref{i-trns}) the Lagrangian (\ref{lagra}) reduces to a new,
simpler one (\ref{lag2}).

We have obtained the new set of field equations
(\ref{EL})-(\ref{E-L2}) for the LRS Bianchi type I spacetime by
using these transformations. We have found the general class of
solutions of BD field equations with potential $U(\Phi)=\lambda
\Phi^2$ in the background of LRS Bianchi type I spacetime
exhibiting Noether symmetry. This solution family contains an
arbitrary function, called $\nu(t)$ in this work, and two
arbitrary constants.

In Section \ref{ornekler}, we gave examples using some simple
forms of the {\em seed function} $\nu(t)$; first  as powers, then
exponentials, powers of $(t_{c}-t)$, and some others. The first
three are chosen to be similar to popular models in FRW cosmology.
The solutions are shown concisely in tables, clearly showing the
relation between the behavior of the model universes and the
parameters of the seed function.

Because of the specialness of the value $W-3/2$ for the BD-theory
when one considers conformal transformations, we considered the
problem also in the {\em Einstein frame} in Section
\ref{einstein}. For this value, the scalar field becomes
non-dynamical, taking on the arbitrary nature of the function
$\nu(t)$ in the {\em Jordan frame}. The solutions found in the
Einstein frame are consistent with those found in the Jordan
frame; they can be transformed into each other.

The models found in Section \ref{ornekler} show a wide range of
behaviors, featuring Big Bangs, Big Crunches, Big Rips, Bounces,
various singularities of the cigar or pancake types, etc. While
some of these solutions are of theoretical interest only, there
are many expanding-universe solutions with acceleration,
consistent with observational data. For example, all the solutions
in Table \ref{tb:l1l2nonzero} and all solutions in subsection
\ref{ustel} except the few strictly contracting solutions, feature
late-time acceleration.

Even the isotropic special case with power-seed
function displays surprising richness, despite the time-dependence
of the scalar field being the same for all powers. One solution is
particularly interesting: It is Minkowski space, containing a
nontrivial scalar field. The particular nonminimal coupling and
the potential selected by the Hessian Determinant condition and
the Noether symmetry make this possible. Given the important
guiding role of the concept of symmetry in modern theoretical
physics, we believe that the family of solutions found and
analyzed in this work constitute a potentially more relevant set
among all possible solutions for LRS Bianchi type-I cosmological
models containing a Brans-Dicke field.

\section*{Acknowledgements}

This work was supported by Akdeniz University, Scientific Research
Projects Unit.


\end{document}